\documentclass[manuscript]{aastex62}

\submitjournal{ApJ}
\usepackage{lineno}
\usepackage{comment}
\usepackage{amsmath}
\usepackage{amssymb}

\begin{document}


\title{Power Anisotropy, Dispersion Signature and Turbulence Diffusion Region in the 3D Wavenumber Domain of Space Plasma Turbulence}

\correspondingauthor{Jiansen He}
\email{jshept@pku.edu.cn}

\author[0000-0001-7655-5000]{Rong Lin}
\affiliation{School of Earth and Space Sciences, Peking University \\
Beijing, 100871, China}

\author[0000-0001-8179-417X]{Jiansen He}
\affiliation{School of Earth and Space Sciences, Peking University \\
Beijing, 100871, China}

\author[0000-0002-1541-6397]{Xingyu Zhu}
\affiliation{School of Earth and Space Sciences, Peking University \\
	Beijing, 100871, China}

\author[0000-0003-2562-0698]{Lei Zhang}
\affiliation{Qian Xuesen Laboratory of Space Technology, China Academy of Space Technology, Beijing, 100094, China}

\author[0000-0002-6300-6800]{Die Duan}
\affiliation{School of Earth and Space Sciences, Peking University \\
	Beijing, 100871, China}

\author{Fouad Sahraoui}
\affiliation{LPP, CNRS, Ecole Polytechnique, Université Paris-Sud, Observatoire de Paris, Université Paris-Saclay, Sorbonne Université, PSL Research University, 91128 Palaiseau, France}

\author[0000-0002-0497-1096]{Daniel Verscharen}
\affiliation{Mullard Space Science Laboratory, University College London, Dorking RH5 6NT, UK}
\affiliation{Space Science Center, University of New Hampshire, Durham NH 03824, USA}

\begin{abstract}
We explore the multi-faceted important features of turbulence (e.g., anisotropy, dispersion, diffusion) in the three-dimensional (3D)  wavenumber domain ($k_\parallel$, $k_{\perp,1}$, $k_{\perp,2}$), by employing the k-filtering technique to the high-quality measurements of fields and particles from the MMS multi-spacecraft constellation. We compute the 3D power spectral densities (PSDs) of magnetic and electric fluctuations (marked as $\rm{PSD}(\delta \mathbf{B}(\mathbf{k}))$ and $\rm{PSD}(\delta \mathbf{E}'_{\langle\mathbf{v}_\mathrm{i}\rangle}(\mathbf{k}))$), both of which show a prominent spectral anisotropy in the sub-ion range. We give the first 3D image of the bifurcation between power spectra of the electric and magnetic fluctuations, by calculating the ratio between $\rm{PSD}(\delta \mathbf{E}'_{ \langle\mathbf{v}_\mathrm{i}\rangle}(\mathbf{k}))$ and $\rm{PSD}(\delta \mathbf{B}(\mathbf{k}))$, the distribution of which is related to the non-linear dispersion relation. We also compute the ratio between electric spectra in different reference frames defined by the ion bulk velocity, that is $\mathrm{PSD}(\delta{\mathbf{E}'_{\mathrm{local}\ \mathbf{v}_\mathrm{i}}})/\mathrm{PSD}(\delta{\mathbf{E}'_{ \langle\mathbf{v}_\mathrm{i}\rangle}})$, to visualize the turbulence ion diffusion region (T-IDR) in wavenumber space. The T-IDR has an anisotropy and a preferential direction of wavevectors, which is generally consistent with the plasma wave theory prediction based on the dominance of kinetic Alfv\'en waves (KAW). This work manifests the worth of the k-filtering technique in diagnosing turbulence comprehensively, especially when the electric field is involved.
\end{abstract}


\section{Introduction}
Plasma turbulence is ubiquitous in most space environments, including the solar wind (SW) \citep{Alexandrova2013,Bruno2013} and the magnetosheath \citep{Sahraoui2003,Huang2014,Sahraoui2020}. The investigation of turbulence is necessary to understand the acceleration, heating and transports of space plasmas \citep{Schekochihin2009,Howes2011}. Turbulence exhibits an anisotropic distribution of fluctuation energy in wavenumber ($\mathbf{k}$) space, i.e., a (power) spectral anisotropy \citep{Horbury2012,Oughton2015,Narita2018}. This kind of anisotropy is typically characterized with weaker and stronger power levels when sampled along the parallel and perpendicular directions relative to the background magnetic field, respectively. In earlier studies, the "Maltese-cross" pattern of the magnetic correlation function at 1 au shows that turbulence exhibits features of a "Slab+2D" configuration \citep{Matthaeus1990}, which consists of parallel waves and perpendicular structures. As an alternative, the Critical Balance (CB) theory proposed by \citet{Goldreich1995} assumes that the magnitudes of the linear (e.g. Alfvén) time scale $\tau_\mathrm{A}$ and of the non-linear time scale $\tau_{\mathrm{NL}}$ are comparable, leading to a balance between the effects of wave propagation and non-linear interaction. Applying the CB theory to Alfvén waves yields scaling indices of power spectral density (PSD): -2 in the parallel direction and -5/3 in the perpendicular direction, which is in agreement with the observed dependence of PSD indices on the sampling direction relative to the local background magnetic field ($\theta_{VB}$) in single-point timeseries analyses of the the SW \citep{Horbury2008}. The observation of the scaling anisotropy at Magnetohydrodynamics (MHD) scales is to some degree sensitive to the determination of the background magnetic field direction. However, the power anisotropy still exists throughout the inertial range \citep{Wu2020}. Anisotropy also exists in the kinetic range \citep{Leamon1998,Sahraoui2010}. Involving appropriate kinetic wave modes in critically balanced cascade models permits the prediction of the anisotropy at kinetic scales. Ideally, a critically balanced Kinetic Alfvén Wave (KAW) cascade would satisfy $k_\|\propto k_\perp^{1/3}$ \citep{Cho_2004,Chen2010}. Spectral anisotropy of turbulence is also found in the magnetosheath \citep{Alexandrova2008, He2011}. A recent statistical and robust survey conducted by \citet{Wang2020} has found a scale-dependent 3D anisotropy pattern down to the electron scales, with the relation $l_{\|} \propto l_{\perp}^{0.72}$ for the parallel and the perpendicular correlation lengths with respect to the background magnetic field in the sub-ion range without being significantly affected by intermittency as previously thought. The observational scaling cannot be explained by the theory based on CB-type KAW, which predicts $l_{\|} \propto l_{\perp}^{1/3}$ (equivalent to $k_\|\propto k_\perp^{1/3}$), and calls for further investigation. The coexistence of different wave modes at kinetic scales (references) complicates the study of small-scale plasma turbulence furthermore \citep{He2011ICW, He2012, Podesta2012, He2015, Zhu_2019}.

Nevertheless, when dealing with data offered by only one spacecraft, as the spatial and temporal variations are entangled, it is impossible to define the full four-dimensional PSD directly. This problem is a key driver for multi-spacecraft missions like Cluster II \citep{Escoubet2001} and Magnetospheric Multiscale Mission (MMS) \citep{Burch2016}, and advanced multi-spacecraft analysis techniques, including the k-filtering technique (also called the wave telescope technique) we apply in this work. It was proposed by \citet{capon1969high}, and introduced to space physics by \citet{pinccon1991local}. As a generalized minimum variance analysis, it gives an estimation of the four-dimensional $\mathrm{PSD}(\omega, \mathbf{k})$ based on Fourier Transformation so that time periodicity and spatial periodicity are separated. Cluster II is the first mission that has offered applicable datasets for the technique \citep{Glassmeier2001, Sahraoui2003}. \citet{Sahraoui2006} investigated an event dominated by mirror-mode turbulence with strong anisotropy, starting the investigation into turbulence with this technique. \citet{Narita2010doppler} presented the first 4D PSD of turbulence, which paved the way for our methodology presented in this article. In this method, the smallest separation between the spacecraft defines the smallest scales of the turbulence that can be resolved. The MMS mission provides high-resolution datasets and smaller satellite separation down to the electron scale for this technique. \citet{Narita2016} presented a comprehensive case study of turbulence based on MMS observations, including wave mode, dispersion relation and propagating direction of waves. \citet{Roberts2019} extended the scope of the measurement beyond the magnetic field, e.g. velocity, density and thermal velocity of ions, and estimated power index anisotropy. However, k-filtering analyses of the turbulence electric field are rare. As an early and inspiring work, \citet{Tjulin2005} argued that in order to use more information and to estimate polarization, we must include the electric field in this technique. After Tjulin, however, we can hardly find applications involving the electric field.

The fluctuating electric field and its PSD play important roles in plasma turbulence. According to Faraday's law, the bifurcation of the PSDs of electric fluctuations and magnetic fluctuations is a signature of dispersion at kinetic scales, which was used by \citet{Bale2005} to support that turbulence has a KAW nature at the sub-ion scale. The electric field is often evoked with the current to estimate the energy-transfer rate between particles and fields as $\mathbf{J}\cdot\mathbf{E}$ or $\mathbf{J}\cdot\mathbf{E}'$ \citep{Zenitani2011,He2019,Duan2020}. Additionally, by computing the electric field in the ion-flow reference frame $\mathbf{E'}=\mathbf{E}+\mathbf{v}_\mathrm{i}\times\mathbf{B}$ and using $|\delta\mathbf{E'}|/|\delta\mathbf{E}|$ as a metric, where $\delta\mathbf{E}$ denotes the fluctuation of the electric field, we determine the plasma demagnetization in magnetic reconnection events \citep{1999PhPl....6.1781H,birn2007reconnection, Lu2010}. This measurement has been adopted by \citet{Duan2018} for the computation of energy transfer in plasma waves based on linear kinetic theory to illustrate the diffusion of magnetic flux relative to the flow of plasma in wavenumber space, which becomes significant when approaching kinetic scales. \citet{2020ApJ...898...43H} further proposed the concept of turbulence ion and electron diffusion regions (T-IDR and T-EDR) based on the observed the ratio $\mathrm{PSD}(\delta {\mathbf{E}'_{\mathrm{i/e, local}}})/\mathrm{PSD}(\delta {\mathbf{E}_{\mathrm{i, global}}})$ as a reasonable alternative to $|\delta\mathbf{E'}|/|\delta\mathbf{E}|$. In conclusion, the spectrum of the fluctuating electric field is related to basic features and mechanisms of plasma turbulence, especially at sub-ion scales. Reliable electric PSDs are necessary for the investigations of this topic.

Out of the considerations above, we apply the k-filtering technique to a case study of plasma turbulence in the magnetosheath, using measurements from MMS. We compute PSDs from the high-resolution magnetic and electric field signals, i.e., $\rm{PSD}(\delta \mathbf{B}(\mathbf{k}))$ and $\rm{PSD}(\delta \mathbf{E}'_{\langle\mathbf{v}_\mathrm{i}\rangle}(\mathbf{k}))$, respectively. We compare these two PSDs to attain the dispersion signature in 3D wavenumber space. We compute the PSD of the electric field in the local ion-flow frame, $\mathrm{PSD}(\delta{\mathbf{E}'_{\mathrm{local}\ \mathbf{v}_\mathrm{i}}})$, and in the global (averaged) ion-flow frame, $\mathrm{PSD}(\delta{\mathbf{E}'_{ \langle\mathbf{v}_\mathrm{i}\rangle}})$, respectively. By calculating the ratio $\mathrm{PSD}(\delta{\mathbf{E}'_{\mathrm{local}\ \mathbf{v}_\mathrm{i}}})/\mathrm{PSD}(\delta{\mathbf{E}'_{ \langle\mathbf{v}_\mathrm{i}\rangle}})$, we identify the T-IDR in 3D wavenumber space. In Section 2, We introduce our dataset and methodology. We present our PSDs of the magnetic and electric fluctuations, and discuss the dispersion signature in Section 3. The T-IDR will be shown and discussed in Section 4.

\section{Methodology and Dataset}
\subsection{The K-filtering Technique and Dealing with the Doppler Effect}

Here we briefly review the k-filtering technique \citep{paschmann2000issi}. We use the original form of the technique, which consists of three main steps. Firstly we apply a Fourier Transformation to the signals, and then combine these resulting Fourier images into a vector $\mathbf{A}(\omega,\mathbf{r}_i)$, e.g. when we choose the magnetic field signals:

\begin{equation}
	\mathbf{A}(\omega, \mathbf{r}_i)=\left[\begin{array}{c}
	A_{x}(\omega, \mathbf{r}_i) \\
	A_{y}(\omega, \mathbf{r}_i) \\
	A_{z}(\omega, \mathbf{r}_i)
	\end{array}\right],
\end{equation}
where the signal $\mathbf{A}$ can be a combination of vector components ($\mathbf{E}$, $\mathbf{B}$ or $[\mathbf{B,E}]^T$), or scalars ($|\mathbf{B}|$, $n$, $T$, etc.). Then we combine the $\mathbf{A}(\omega,\mathbf{r}_i)$ from the four different satellites together to $[\mathbf{A}(\omega,\mathbf{r}_1),\mathbf{A}(\omega,\mathbf{r}_2),\mathbf{A}(\omega,\mathbf{r}_3),\mathbf{A}(\omega,\mathbf{r}_4)]^T$(abbreviated as $\mathbf{A}(\omega)$ with $\mathbf{r}_i$ omitted), and with proper averaging, we have the covariance matrix $\mathsf{M}_{A}(\omega)$,

\begin{equation}
	\mathsf{M}_{A}(\omega)=E\left[\mathbf{A}(\omega) \mathbf{A}^{\dagger}(\omega)\right].
\end{equation}
We introduce the steering (wave-propagating) matrix $\mathsf{H}(\mathbf{k})$, where $\mathsf{I}$ denotes the unit matrix,

\begin{equation}
	\mathsf{H}(\mathbf{k})=\left[\begin{array}{c}
\mathsf{I}\mathrm{e}^{-\mathrm{i} \mathbf{k} \cdot \mathbf{r}_{1}} \\
\mathsf{I}\mathrm{e}^{-\mathrm{i} \mathbf{k} \cdot \mathbf{r}_{2}} \\
\vdots \\
\mathsf{I}\mathrm{e}^{-\mathrm{i} \mathbf{k} \cdot \mathbf{r}_{N}}
\end{array}\right].
\end{equation}
Using the Lagrange multiplier technique, we obtain for the estimation of PSD the following expression,

\begin{equation}
\mathrm{PSD}(\omega, \mathbf{k})=\operatorname{Tr}\left\{\left[\mathsf{H}^{\dagger}(\mathbf{k}) \mathsf{M}_{A}^{-1}(\omega) \mathsf{H}(\mathbf{k})\right]^{-1}\right\}.
\end{equation}

Previous studies remind us of several limitations of this technique \citep{Sahraoui2010limitation}.The technique is based on the assumption of quasi-stationarity and quasi-homogeneity of the turbulent fluctuations. Moreover, the geometric configuration of the spacecraft constellation determines the first Brillouin zone where the estimation is reliable. Due to the spatial aliasing effect that fluctuations separated by $\Delta k=2n\pi/d$ in k-space cannot be distinguished. The maximum wavenumber  $k_{\max}$ resolved in this technique is $\pi/d$, where the $d$ is the appropriate (averaged) separation of spacecraft in the constellation. This wavenumber $k_{\max}$ corresponds to a wavelength of $\lambda=2d$. A configuration of the constellation close to a regular tetrahedron maximizes the Brillouin zone volume and reduces angular aliasing, which would otherwise create fake anisotropy \citep{Narita2009}. Empirically given by \citet{Sahraoui2010limitation} to guarantee accurate positions of power peaks, the minimum resolved wavenumber $k_{\min}$ corresponds to $\sim \pi/5d$. 

Due to the Doppler effect, there is a shift between the frequency of fluctuations measured in the spacecraft frame ($\omega_{\mathrm{sc}}$) and the frequency in the reference frame defined by the plasma flow ($\omega_{\mathrm{pl}}$). In order to compare with theoretical predictions and eliminate the Doppler shift, we reconstruct $\mathrm{PSD}(\omega_\mathrm{pl},\mathbf{k})$ from the direct result $\mathrm{PSD}(\omega_{\mathrm{sc}},\mathbf{k})$. Using the Doppler relation $\omega_\mathrm{pl}=\omega_\mathrm{sc}-\mathbf{k} \cdot \mathbf{v}$, we map the value of $\mathrm{PSD}(\omega_{\mathrm{sc}},\mathbf{k})$ directly to the corresponding $\mathrm{PSD}(\omega_\mathrm{pl},\mathbf{k})$. As the relation $\omega_\mathrm{pl}=\omega_\mathrm{sc}-\mathbf{k} \cdot \mathbf{v}$ corresponds to oblique planes in the 4D ($\omega_\mathrm{sc},\mathbf{k}$) space if $\omega_\mathrm{pl}$ is set, the reconstruction method can be symbolized like cutting 4D bread into pieces with an oblique knife. In the computation, independent variables are discrete, so linear interpolation is employed here. When $\omega_\mathrm{pl}$ is linked with negative $\omega_\mathrm{sc}$, the equality $\mathrm{PSD}(\omega,\mathbf{k})=\mathrm{PSD}(-\omega,-\mathbf{k})$ is used. A figuratively description of this procedure is included in the Appendix.

\subsection{Dataset for Analysis}

\begin{figure}[htb!]
	\centering
	\includegraphics[width=15cm,clip=]{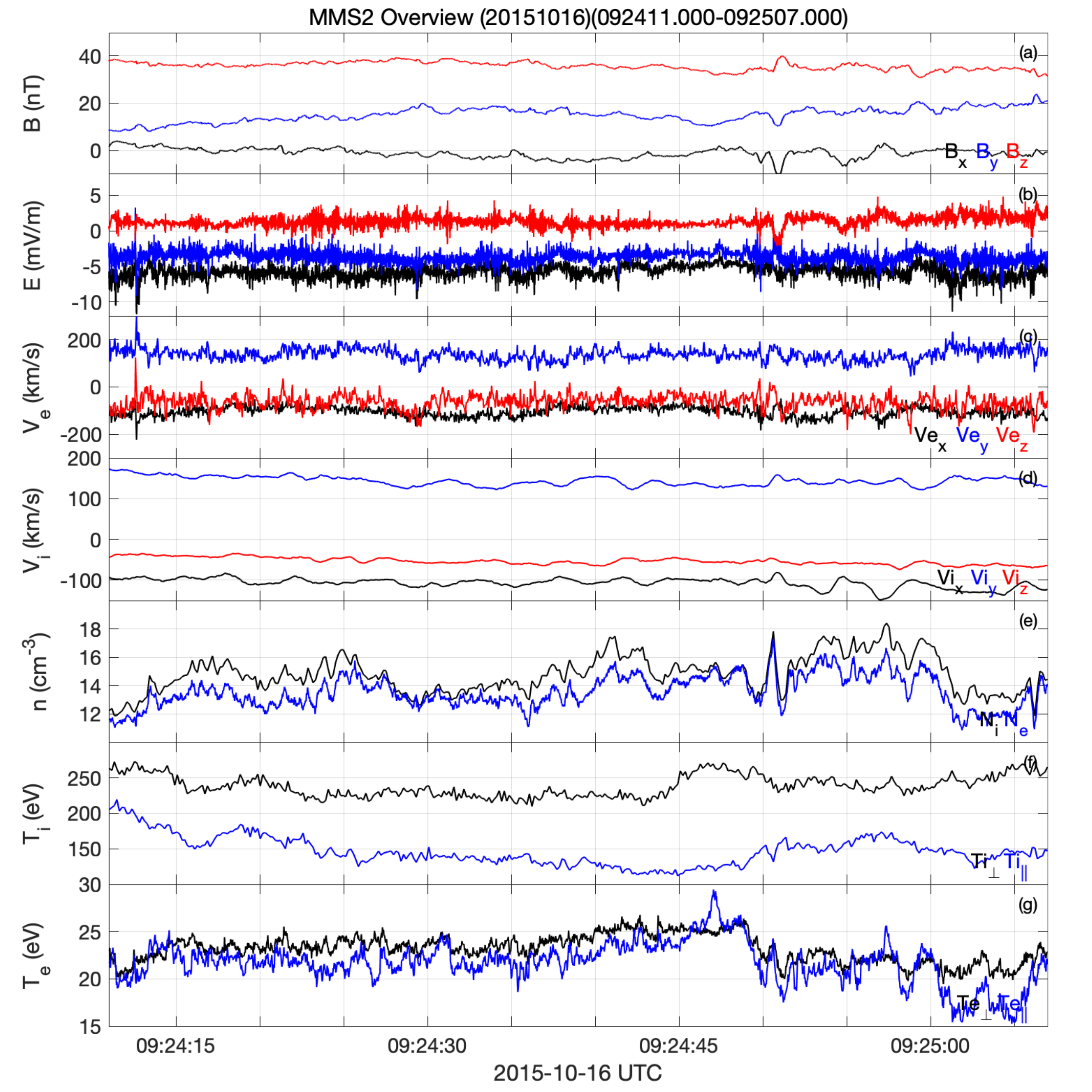}
	\caption{Time series of the MMS2 observations in geocentric solar ecliptic (GSE) coordinates, including (a)  the magnetic field, (b) the electric field, (c, d)the velocities of the electrons and ions, (e) the ion number density, (f, g)the temperature of the ions and electrons. During the time interval, the direction of background magnetic field and bulk flow of particles stayed quasi-stationary.}
	\label{fig1}
\end{figure}

The observation analysed is from MMS from 9:24:11 to 9:25:07, October 16th, 2015, when the satellites are located in the magnetosheath close to the dusk magnetopause, at a distance of $11.9R_\mathrm{E}$ from the earth. The same event has been reported by \citet{Chen2017}, as an event with anisotropy and kinetic Alfvén nature in the sub-ion range, but we neglect the last 17 seconds because the background magnetic field $\mathbf{B}_0$ rotated by about 10 degrees then. The magnetic field data is from the FGM instrument \citep{Russel2016}. The electric field is measured by the SDP \citep{Lindqvist2016} and ADP \citep{Ergun2016} instruments. We use data from the FPI instrument for density, bulk velocity and temperature of the protons \citep{Pollock2016}. The timeseries of the event is shown in Figure \ref{fig1}, with all the variables in the Geocentric Solar Ecliptic (GSE) coordinate system. Some key parameters are listed in Table \ref{table1}. The Alfvén speed exceeds the average ion flow velocity, which potentially causes a violation of the assumptions underlying Taylor's hypothesis. The k-filtering technique does not depend on Taylor's hypothesis, though.

\begin{table}[htb!]
	\centering
	\caption{Key parameters of the event.}
	\begin{tabular}{cc}
		               \textbf{Variable Name}                &        \textbf{Value}         \\ \hline
		 $B_0$, magnitude of the background magnetic field   &            38.8 nT            \\
		       $V_\mathrm{0,i}$, average ion velocity        & 187 $\mathrm{km\cdot s^{-1}}$ \\
		            $V_\mathrm{A}$, Alfvén speed             & 229 $\mathrm{km\cdot s^{-1}}$ \\
		       $n_\mathrm{i}$, number density of ions        &    14.9 $\mathrm{cm^{-3}}$    \\
		    $n_\mathrm{e}$, number density of electrons      &    13.3 $\mathrm{cm^{-3}}$    \\
		        $T_\mathrm{i}$, Temperature of ions          &            208 eV             \\
		      $T_\mathrm{e}$, Temperature of electrons       &             23 eV             \\
		      $\beta_\mathrm{i}$, plasma beta of ions        &             0.86              \\
		    $\beta_\mathrm{e}$, plasma beta of electrons     &             0.096             \\
		         $d_\mathrm{i}$, ion inertial range          &             59 km             \\
		     $\rho_\mathrm{i}$, ion thermal gyroradius       &             54 km             \\
		$d_{\mathrm{avg}}$, average separation of satellites &            14.4 km            \\ \hline
	\end{tabular}
	\label{table1}
\end{table}

\section{3D Power Spectral Densities: Spectral Anisotropy and Dispersion Signature}

\begin{figure}[htb!]
	\centering
	\includegraphics[width=18cm]{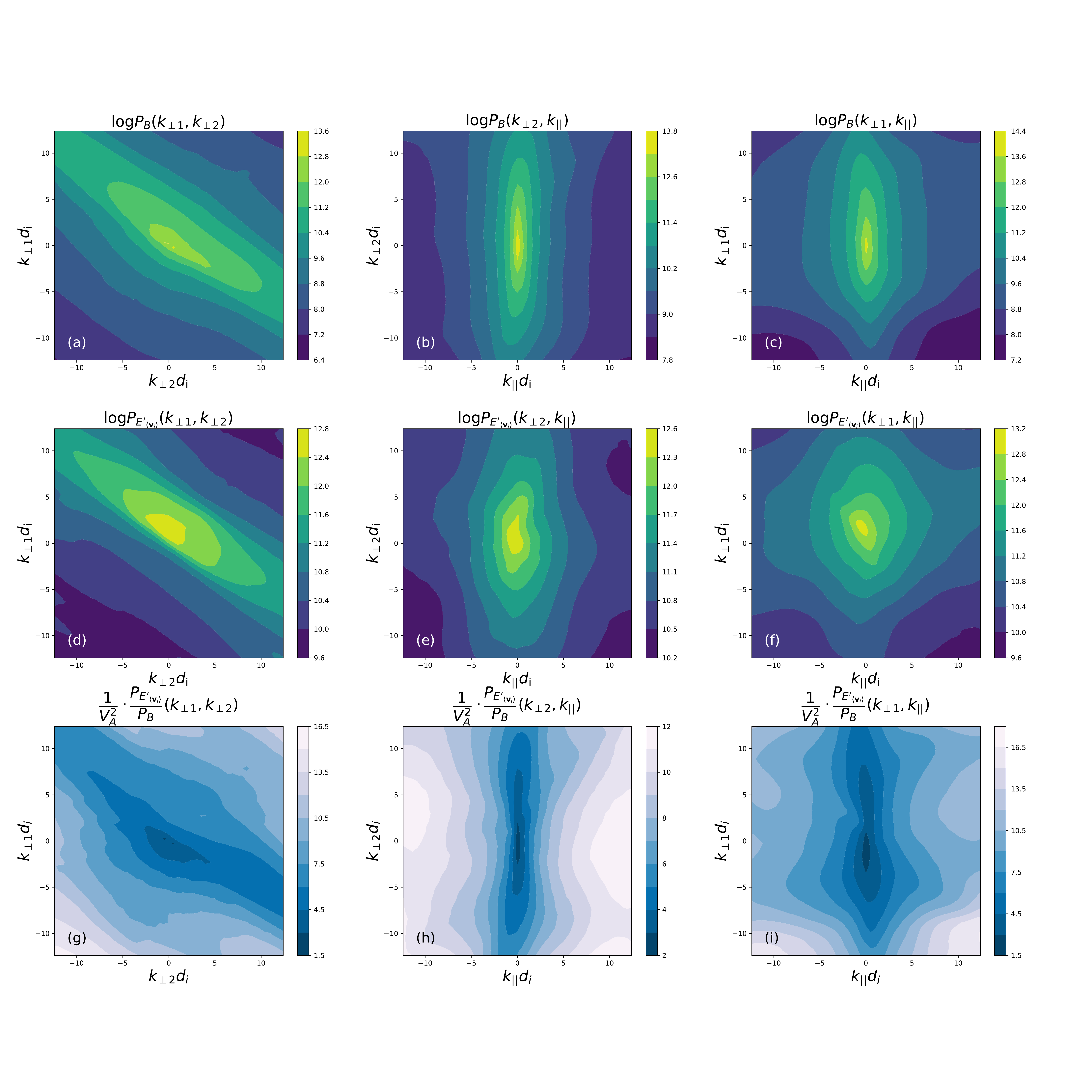}
	\caption{(a-f): Reduced 2D distributions of $P_{\mathbf B}(\omega_\mathrm{pl},\mathbf{k})$ and $P_{\rm{\mathbf{E}}'_{\langle \rm{\mathbf{v}_i}\rangle}}(\omega_\mathrm{pl},\mathbf{k})$, where $P(k_\|,k_{\perp1})=\displaystyle{\iint}P(\omega_\mathrm{pl},\mathbf{k})\mathrm{d}k_{\perp2}\rm{d}\omega_{\rm{pl}}$, and so forth. (g-i): $P_{\rm{\mathbf{E}}'_{\langle \rm{\mathbf{v}_i}\rangle}}/P_\mathbf B$, normalized with the square of the Alfvén speed, $V_\mathrm{A}^2s$, plotted to present the detail of the spectral bifurcation between $P_\mathbf B$ and $P_{\rm{\mathbf{E}}'_{\langle \rm{\mathbf{v}_i}\rangle}}$.}
	\label{fig2}
\end{figure}

The analysed time interval fulfills the conditions for the application of the k-filtering method. The satellites almost formed a regular tetrahedron with $Elongation=0.09$ and $Planarity=0.17$ \citep{robert1998tetrahedron}. The average separation of the satellites, $d_{\mathrm{avg}}$, sets a $k_{\max}\sim0.21 \mathrm{km^{-1}} $, consistent with $k_{\max}d_\mathrm{i}\sim12$, and $k_{\min} \sim 0.04\mathrm{km^{-1}}$, consistent with $k_{\min}d_\mathrm{i}\sim 2.4$. We rotate the reference frame so that $\mathbf{B}_0 =B_0 \hat{e}_\parallel$, and set $\hat{e}_{\perp1}$ and $\hat{e}_{\perp2}$ along $x_{\mathrm{GSE}}\times\mathbf{B}_0$ and $\mathbf{B}_0\times\hat{e}_{\perp1}$, respectively. Figure \ref{fig2}(a-f) show the reduced 2D PSDs of the magnetic and the electric fields, $\mathrm{PSD}(\delta \mathbf{B})$ and $\mathrm{PSD}(\delta {\mathbf{E}}'_{\langle\mathbf{v}_\mathrm{i}\rangle})$, abbreviated as $P_{\rm\mathbf{B}}$ and $P_{\rm{\mathbf{E}}'_{\langle \rm{\mathbf{v}_i}\rangle}}$ henceforth. To reduce the 4D spectrum to 2D in the ($k_i, k_j$) plane, we integrate the 4D PSD over $\omega_\mathrm{pl}$ and the other component of $\mathbf{k}$. In these panels, we find a spectral anisotropy in the wavenumber space for both the magnetic and electric fluctuations, although the electric fluctuations are less anisotropic. Figure \ref{fig2} gives the first 3D image of the bifurcation between $P_{\rm\mathbf{B}}$ and $P_{\rm{\mathbf{E}}'_{\langle \rm{\mathbf{v}_i}\rangle}}$. 

Panels (a) and (d) give an indication of a non-axisymmetry in the power distribution, showing elongated power distributions. This pattern of $P(k_{\perp1}, k_{\perp2})$ reflects that in this event with a relatively short time interval, the kinetic waves have limited directions of $\mathbf{k}_\perp$, which do not distribute evenly in the whole range of azimuthal angle. We compute the $P_{\rm{\mathbf{E}}'_{\langle \rm{\mathbf{v}_i}\rangle}}/P_{\rm\mathbf{B}}$ ratio and normalize it to the Alfvén speed, which is presented in Figure \ref{fig2}(g-i). The ratio is well below the unity at small $\mathbf{k}$ (large scales), and increases to values greater than unity when $|\mathbf{k}|$ increases along the direction perpendicular to $\mathbf{B}_0$, where the fluctuation energy concentrates. The ratio increases even more steeply along the directions not perpendicular to $\mathbf{B}_0$. It exceeds 10 at $|\mathbf{k}|=10$ when $k_\|\gg k_\perp$.

\begin{figure}[htb!]
	\centering
	\includegraphics[width=20cm]{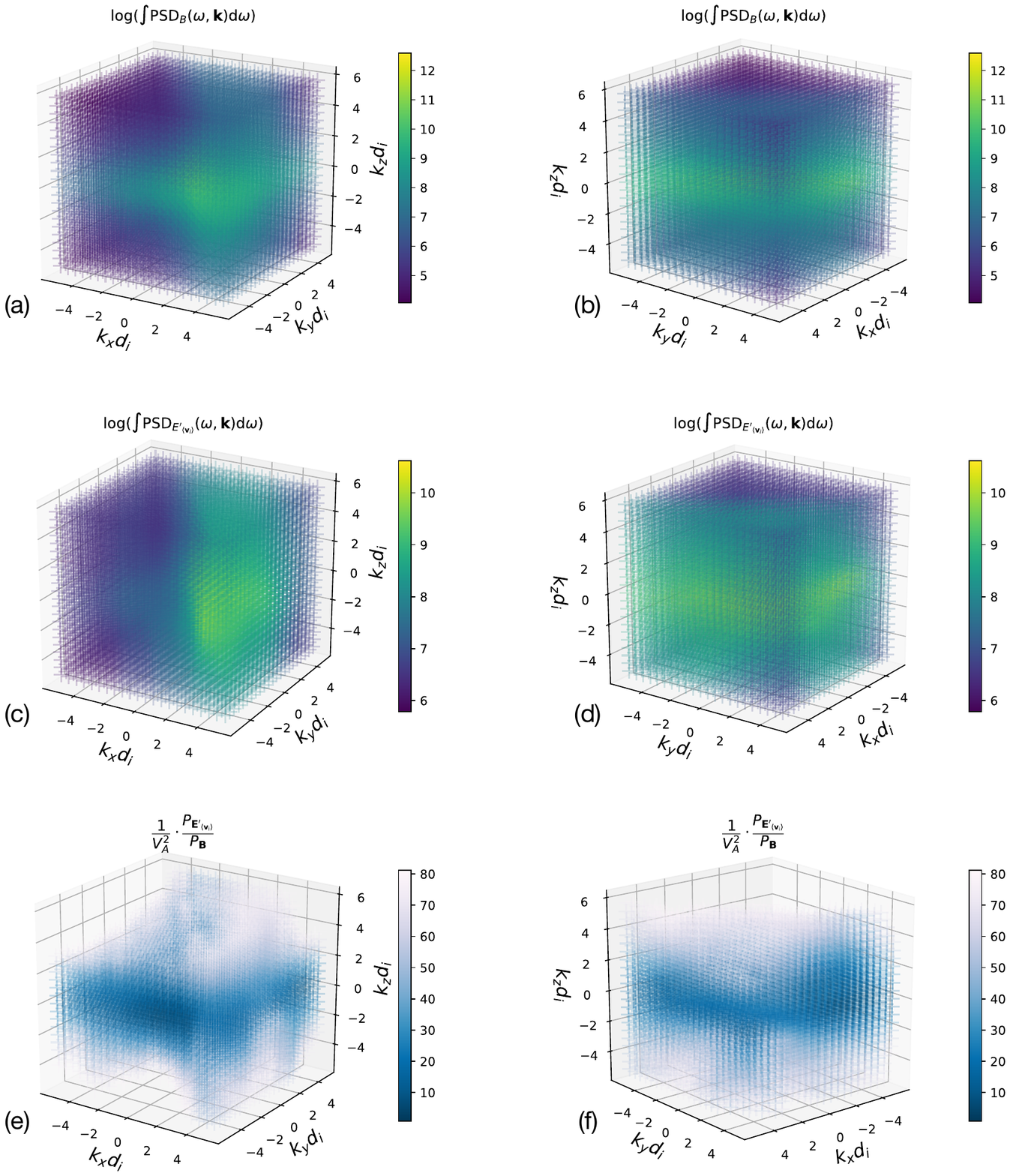}
	\caption{A 3D version of the distributions from Figure \ref{fig2}, where P denotes the reduced spectral distribution integrated over $\omega_{\mathrm{pl}}$ from $\mathrm{PSD}(\omega_{\mathrm{pl}},\mathbf{k})$.}
	\label{fig3}
\end{figure}

In Figure \ref{fig3}, we present $P_{\rm\mathbf{B}}$, $P_{\rm{\mathbf{E}}'_{\langle \rm{\mathbf{v}_i}\rangle}}$ and $\dfrac{1}{V_\mathrm{A}^2}\dfrac{P_{\rm{\mathbf{E}}'_{\langle \rm{\mathbf{v}_i}\rangle}}}{P_{\rm\mathbf{B}}}$ in 3D wavenumber space. The reduced PSDs in wavenumber space are shown after integration over $\omega_{\mathrm{pl}}$. All properties mentioned in the above 2D plots are also visible in these 3D plots.

\section{The Turbulence Ion Diffusion Region in Wavenumber Space}

\begin{figure}[htb!]
	\centering
	\includegraphics[width=18cm]{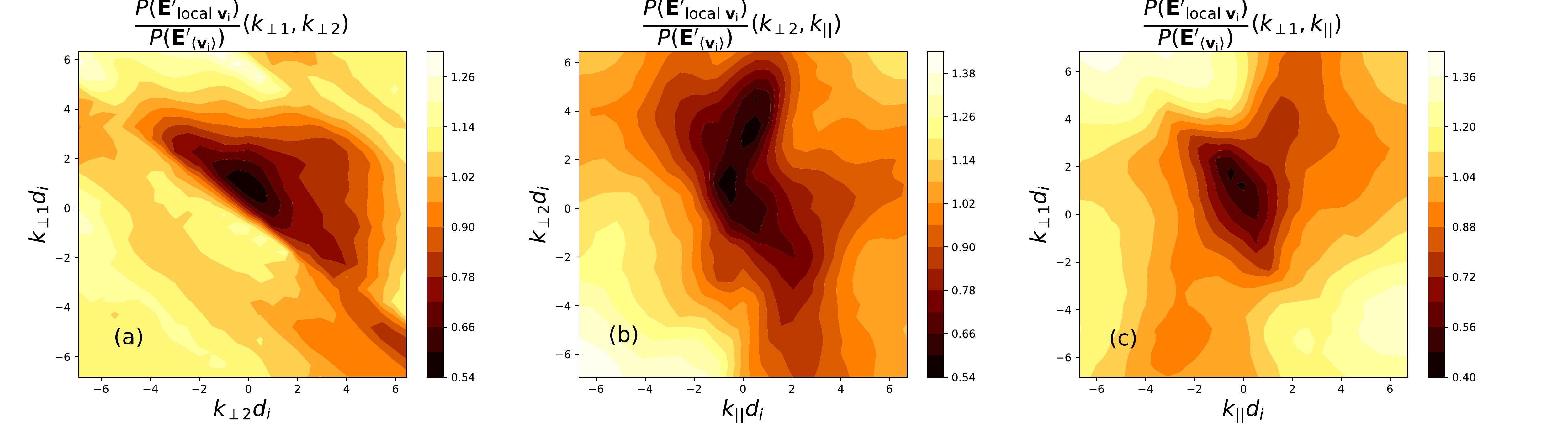}
	\caption{$P_{\rm{\mathbf{E}}'_{\rm{local\ \mathbf{v}_i}}}/P_{\rm{\mathbf{E}}'_{\langle \rm{\mathbf{v}_i}\rangle}}$ of the event, where $P$ denotes PSD integrated over $\omega_{\mathrm{pl}}$ and different components of $\mathbf{k}$ as defined in Figure \ref{fig2}.
	$\displaystyle\frac{P_{\rm{\mathbf{E}}'_{\rm{local\ \mathbf{v}_i}}}}{P_{\rm{\mathbf{E}}'_{\langle \rm{\mathbf{v}_i}\rangle}}}(k_\|,k_{\perp1})=\left({\iint}P_{\rm{\mathbf{E}}'_{\rm{local\ \mathbf{v}_i}}}(\omega_\mathrm{pl},\mathbf{k})\mathrm{d}k_{\perp2}\rm{d}\omega_{\rm{pl}}\right)/\left({\iint}P_{\rm{\mathbf{E}}'_{\langle \rm{\mathbf{v}_i}\rangle}}(\omega_\mathrm{pl},\mathbf{k})\mathrm{d}k_{\perp2}\rm{d}\omega_{\rm{pl}}\right)$.}
	\label{fig4}
\end{figure}

\begin{figure}[htb!]
	\centering
	\includegraphics[width=14cm]{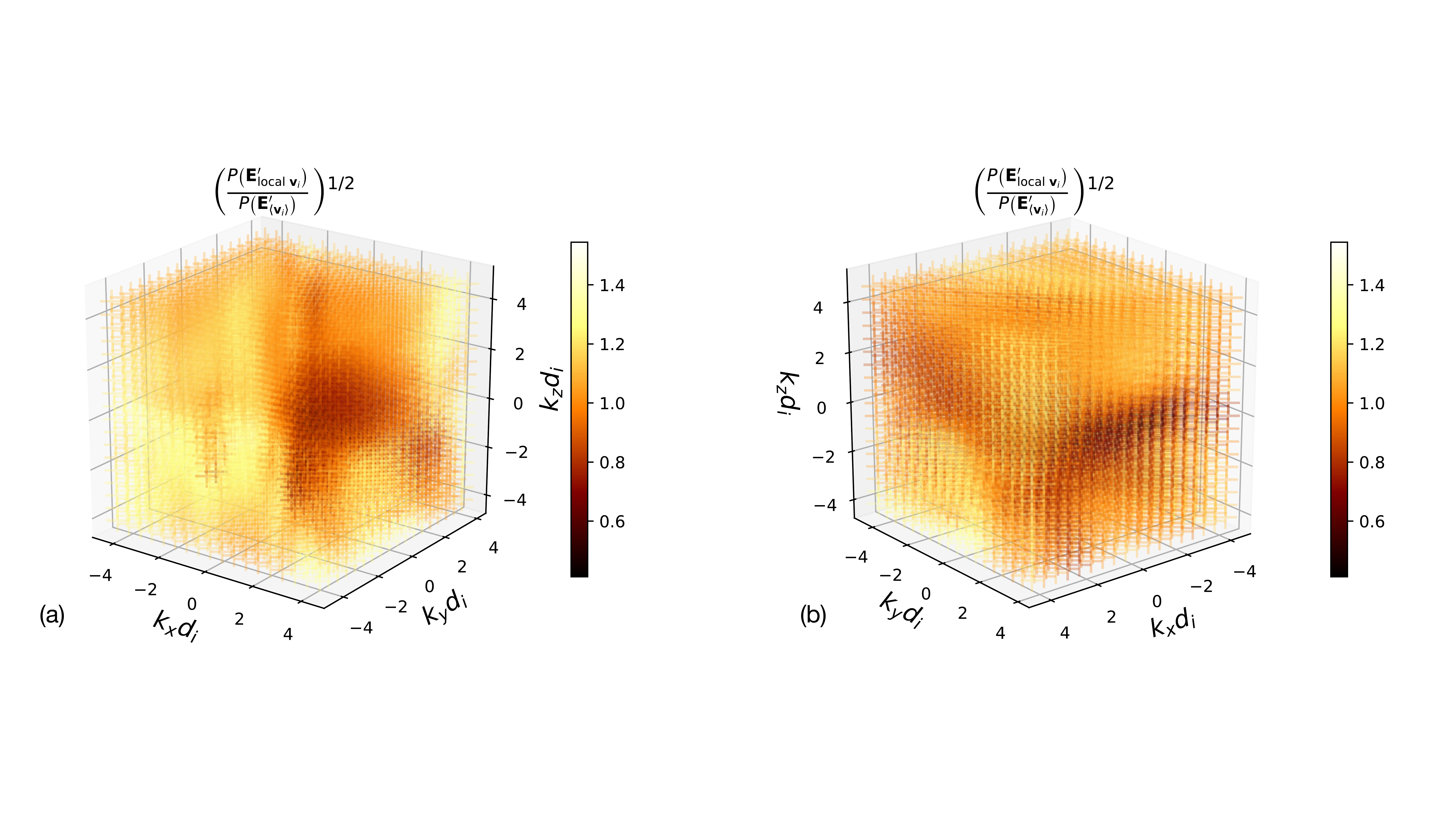}
	\caption{Reduced 3D distribution of the ratio $P_{\rm{\mathbf{E}}'_{\rm{local\ \mathbf{v}_i}}}/P_{\rm{\mathbf{E}}'_{\langle \rm{\mathbf{v}_i}\rangle}}$ after integration over $\omega_\mathrm{pl}$.}
	\label{fig5}
\end{figure}

We further compute the ratio of $P_{\rm{\mathbf{E}}'_{\rm{local\ \mathbf{v}_i}}}$ to $P_{\rm{\mathbf{E}}'_{\langle \rm{\mathbf{v}_i}\rangle}}$ to measure the demagnetization and the diffusion of ions. These two PSDs are computed from electric fields in the local ion-flow frame, $\mathbf{E}'_{\rm{local }\ \mathbf{v}_\mathrm{i}}=\mathbf{E}_\mathrm{sc}+\mathbf{v}_\mathrm{i}\times\mathbf{B}$, and in the global ion-flow frame, $\mathbf{E}'_{\langle\mathbf{v}_\mathrm{i}\rangle}=\mathbf{E}_\mathrm{sc}+\langle\mathbf{v}_\mathrm{i}\rangle\times\mathbf{B}$, where the angle brackets mean averaging over the whole interval, respectively. Figure \ref{fig4} shows the ratio after appropriate integrations. As the figure shows, the ratio is lower in regions close to quasi-perpendicular wavevectors ($\mathbf{k}_\perp$) than near quasi-parallel wavevectors ($k_\parallel$). In panel (a), the region with a low ratio (less than 1) covers a large range of the parameter space. In panel (b), the low ratio range reaches $k_{\perp2}d_\mathrm{i}=5$, and unlike the PSD, it is far less symmetric in $k_{\perp2}$. The corresponding 3D version is shown in Figure \ref{fig5}.

\begin{figure}[htb!]
	\centering
	\includegraphics[width=18cm,clip=]{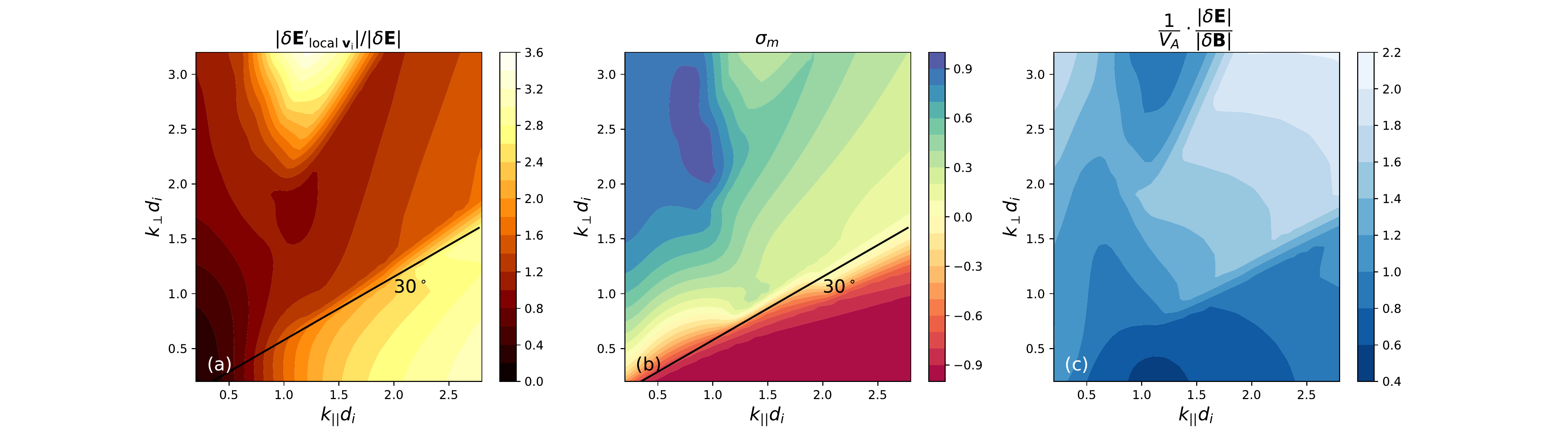}
	\caption{Metric of ion demagnetization/diffusion (left), normalized magnetic helicity (middle), and metric of dispersion computed by NHDS. The black line denotes $\theta_{kB}\sim30^\circ$.}
	\label{fig6}
\end{figure}

Using the New Hampshire Dispersion Relation Solver (NHDS) \citep{Verscharen2018NHDS} based on linear Vlasov-Maxwell theory, we calculate the fluctuations $\delta\mathbf{B}$, $\delta\mathbf{E}$, $
\delta\mathbf{v}_\mathrm{i}$ and $\delta\mathbf{v}_\mathrm{e}$ of eigenmodes belonging to the Alfvén-mode branch under the same plasma condition as measured (see Table \ref{table1}). 
Figure \ref{fig6} is derived from the output of NHDS. Figure \ref{fig6}(a) shows the modelled distribution of $|\delta \mathbf{E}'_{\mathrm{local}\ \mathbf{v}_\mathrm{i}}|/|\delta \mathbf{E}|$ in wavenumber space in the plasma frame, which is considered equivalent to the square root of $P_{\rm{\mathbf{E}}'_{\rm{local~\mathbf{v}_i}}} / P_{\rm{\mathbf{E}}'_{\langle \rm{\mathbf{v}_i}\rangle}}$. The ratio $|\delta \mathbf{E}'_{\mathrm{local}\ \mathbf{v}_\mathrm{i}}|/|\delta \mathbf{E}|$ stays below 1.0 when $k_\| d_\mathrm{i}<0.5$ and $k_\perp d_\mathrm{i}<1.0$. It exceeds unity at smaller scales. A sharp edge near $\theta_{kB}=30^\circ$ is visible in panel (a), which separates the high and low ratios at small and large $\theta_{kB}$, respectively. Another ridge-like boundary between low and high ratios is also found at $\theta_{kB}=70^\circ$ when $kd_\mathrm{i}\gtrsim2.5$. Figure \ref{fig6}(b) displays the normalized magnetic helicity $\sigma_m$. We also see an edge at $\theta_{kB}\sim30^\circ$ in this panel corresponding to $\sigma_m=0$. The normalized magnetic helicity $\sigma_m$ is negative (positive) at $\theta_{kB}<30^\circ$ ($\theta_{kB}>30^\circ$). The positive $\sigma_m$ at large $\theta_{kB}$ is an evident feature of KAW, which is distinguished from the negative $\sigma_m$ at small $\theta_{kB}$ characteristic for ion cyclotron waves (ICWs). 

According to Figure \ref{fig6}(a), when ICWs dominate, the metric of diffusion, $P_{\rm{\mathbf{E}}'_{\rm{local\ \mathbf{v}_i}}} / P_{\rm{\mathbf{E}}'_{\langle \rm{\mathbf{v}_i}\rangle}}$, exceeds 1.0 significantly. 
For the ICWs at ion scales, the fluctuating electric field $\delta{\rm{\mathbf{E}}}$ can be approximated as $-\delta{\rm{\mathbf{v}}}_{{\rm e}}\times{\rm{\mathbf{B}}}_{{0}}$, since the electrons are still frozen-in with the magnetic  fields. Therefore, the electric field in the local ion-flow frame can be approximated as $\delta{\rm{\mathbf{E}}}'\sim(-\delta{\rm{\mathbf{v}}}_{{\rm e}}+\delta{\rm{\mathbf{v}}}_{{\rm i}})\times{\rm\mathbf{B}}_{{0}}$. On the other hand, we know that $|\delta{\rm{\mathbf{v}}}_{\rm i}|\textgreater  |\delta{\rm{\mathbf{v}}}_{\rm e}|$ and $|\delta{\rm{\mathbf{v}}}_{\rm i}-\delta{\rm{\mathbf{v}}}_{\rm e}|\textgreater |\delta{\rm{\mathbf{v}}}_{\rm e}|$ for ICWs, which means that ions are the primary current carrier of the wave-current density, according to the plasma wave theory. Therefore, we anticipate that $|\delta{\rm{\mathbf{E}}}'|\textgreater |\delta{\rm{\mathbf{E}}}|$ for the ICWs. Comparing to this theoretical prediction, we observe that the k-filtering results do not show such a high ratio of the ion diffusion metric at quasi-parallel wavenumbers. This finding suggests that ICWs are not detectable in our measurement interval.

Figure \ref{fig6}(c) is the modelled distribution of $|\delta \mathbf{E}|/|\delta\mathbf{B}|$, normalized to the Alfvén speed $V_\mathrm{A}$. It can be compared with Figure \ref{fig2}(g-i) and Figure \ref{fig3}(e)\&(f). They are generally consistent at quasi-perpendicular $\mathbf{k}$s, but this similarity diminishes at small $\theta_{kB}$, where the theoretical distribution corresponding to ICWs does not find a counterpart in the k-filtering result. As the polarization (i.e. the imaginary part of the complex components) of the electric fields cannot be determined by k-filtering, we cannot tell exactly what the full wave-mode composition of our measurement interval is and cannot explain more about the distribution. It is possible that waves with a strong electrostatic component (e.g., quasi-parallel ion acoustic waves) may contribute to the corresponding power distribution. Nevertheless, our comparison of the dispersion properties measured by $|\delta \mathbf{E}|/|\delta\mathbf{B}|$ reveals the dominant modes in our time interval as quasi-perpendicular kinetic Alfvén waves.

\section{Summary and discussion}

For a case study of turbulence in the terrestrial magnetosheath, we compute the 4D PSDs of the turbulent magnetic and electric fields, recalculate the frequency in the plasma frame from its counterpart in the spacecraft frame, and reconstruct the PSDs with the frequency in the plasma frame. Both $P_{\rm\mathbf{B}}$ and $P_{\rm{\mathbf{E}}'_{\langle \rm{\mathbf{v}_i}\rangle}}$ show the typical anisotropy in the 2D plane of ($k_\|,k_\perp$), but $P_{\rm{\mathbf{E}}'_{\langle \rm{\mathbf{v}_i}\rangle}}$ is less anisotropic. We also find some degree of non-axisymmetry in the ($k_{\perp1},k_{\perp2}$) plane, which agrees with \citet{Roberts2019}. Such non-axisymmetry with some elongation of the PSD suggests that oblique kinetic Alfvén waves at various $\mathbf{k}_\perp$ are concentrated within a finite range of angles instead of the $2\pi$ omni-direction.

The bifurcation between $P_{\rm\mathbf{B}}$ and $P_{\rm{\mathbf{E}}'_{\langle \rm{\mathbf{v}_i}\rangle}}$ was evaluated through $\dfrac{1}{V_\mathrm{A}^2}\dfrac{P_{{\mathbf E}'_{\langle \mathbf{v}_\mathrm{i} \rangle}}}{P_\mathbf B}$ usually considered as a indicator for wave dispersion in turbulence. We found in the quasi-perpendicular direction where the fluctuation energy concentrates, that the bifurcation agrees with the prediction for KAWs according to linear Vlasov-Maxwell theory. Apart from these features, the observation of $\dfrac{1}{V_\mathrm{A}^2}\dfrac{P_{{\mathbf E}'_{\langle \mathbf{v}_\mathrm{i} \rangle}}}{P_\mathbf B}$ cannot be solely explained by the theoretical prediction only based on the Alfvén wave branch.

For the first time, we computed the ratio between $\mathrm{PSD}(\delta{\mathbf{E}'_{\mathrm{local}\ \mathbf{v}_\mathrm{i}}})$ in the local ion bulk flow frame and $\mathrm{PSD}(\delta{\mathbf{E}'_{ \langle\mathbf{v}_\mathrm{i}\rangle}})$ in the global ion bulk flow frame. We computed the electric field power ratio to identify the ion demagnetization effect in wavenumber space at kinetic scales and compared it with the prediction of ion demagnetization in the Alfv\'en wave branch as calculated by NHDS. The ratio of the event exhibits anisotropy and asymmetry of wavevectors. Considering the strengths of this method as described above, we anticipate that the k-filtering technique will become a powerful technique for the comprehensive investigation of diffusion physics in turbulence.

Although the k-filtering technique has shown considerable capability, it suffers from limitations and errors, as mentioned above. Especially spatial aliasing is a central issue, the challenge of which has been theoretically discussed by \citet{Narita2009}. The spatial aliasing effect prevents us from quantifying the distribution of PSDs in ($\omega,\mathbf{k}$) and quantitatively comparing reduced PSDs with those obtained directly from Fourier/wavelet transformations of the time sequences. The limitation of this method to around (or even less than) one decade in wavevector space is another weakness of this technique, which prevents the simultaneous analysis of different plasma scales. A more flexible constellation with more than 4 satellites would be a prospect to be anticipated \citep{zhang2020conditions, Dai2020}. Unfortunately, phase spectra of variables in ($\omega$, $\mathbf{k}$) space cannot be acquired through the k-filtering technique. As a consequence, we can not further study the power spectral distribution of different component disturbances (e.g., $\delta E_\|$ and $\delta E_\perp$), nor can we study the combined spectra of different disturbances (e.g., $\delta {\rm{\mathbf{J}}}\cdot \delta{\rm{\mathbf{E}}}$ for the dissipation rate spectrum).

\bigbreak

\noindent Acknowledgements:

The authors are grateful to the teams of the MMS spacecraft for providing the data. The authors from China are supported by NSFC under contracts 41874200, 41421003, and by CNSA under D020301 and D020302. D.V. is supported by the STFC Ernest Rutherford Fellowship ST/P003826/1 and STFC Consolidated Grant ST/S000240/1.

We sincerely thank Olga Alexandrova for the discussion with us at the vEGU meeting 2021.

\begin{figure}[htb!]
	\centering
	\includegraphics[width=18cm,clip=]{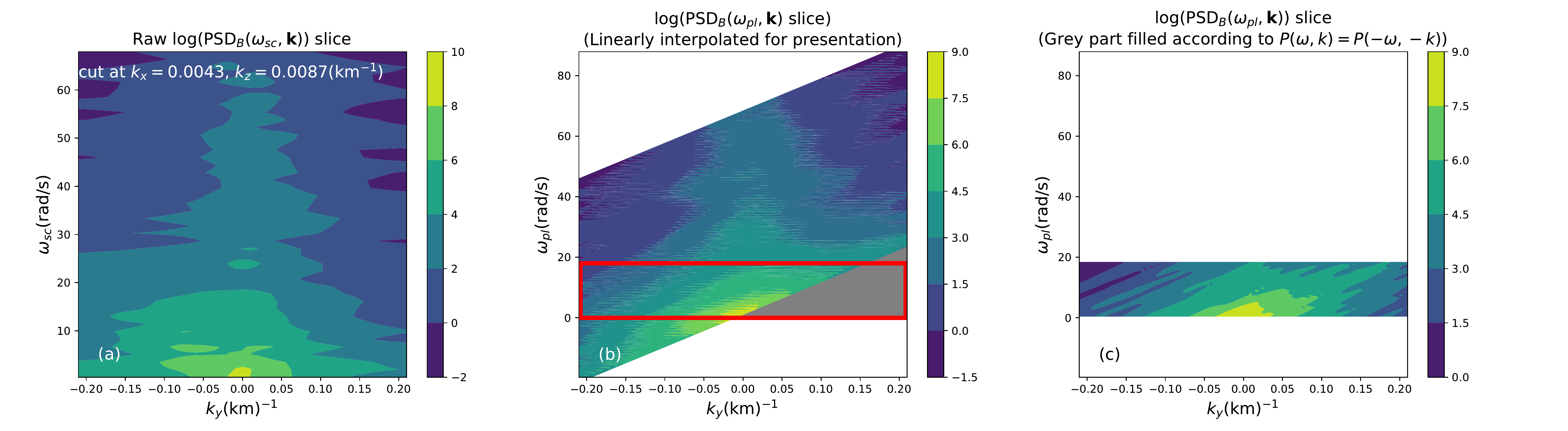}
	\caption{Cuts of energy distribution in the $\omega-k$ plane. The angular frequencies are: (a) in the spacecraft frame, (b) in the plasma frame obtained by the Doppler relation. Panel(c) shows a result after filling the grey part in panel(b) with its mirror counterpart.}
	\label{fig7}
\end{figure}

\appendix
We described our method with Figure \ref{fig7}, which imitating the presentation of Figure 9 in the paper \citet{Narita2010MagneticEnergy}. Panel(a) is a slice cut from the $\mathrm{PSD}_B(\omega_{\mathrm{sc}},\mathbf{k})$ at $k_x=0.0043 \mathrm{km^{-1}}$ and $k_z=0.0087 \mathrm{km^{-1}}$. After the Doppler shift correction, the slice’s shape changes from a rectangle to a parallelogram, which is shown in panel(b). For a proper integration following, we require a box-shaped $\mathrm{PSD}_B(\omega_{\mathrm{pl}},\mathbf{k})$, so we reserve the distribution in the red rectangle in panel(b) and append the grey patch from its mirror counterpart using the equality mentioned above. Panel(c) shows the resulted slice with frequencies in the plasma frame. The 4D $\mathrm{PSD}_B(\omega_{\mathrm{pl}},\mathbf{k})$ is the ensemble of all the slices like this one.

\bibliographystyle{aasjournal}
\bibliography{references}

\begin{thebibliography}{}
\expandafter\ifx\csname natexlab\endcsname\relax\def\natexlab#1{#1}\fi

\bibitem[{Alexandrova {et~al.}(2013)Alexandrova, Chen, Sorriso-Valvo, Horbury,
  \& Bale}]{Alexandrova2013}
Alexandrova, O., Chen, C.~H., Sorriso-Valvo, L., Horbury, T.~S., \& Bale, S.~D.
  2013, Space Science Reviews, 178, 101

\bibitem[{Alexandrova {et~al.}(2008)Alexandrova, Lacombe, \&
  Mangeney}]{Alexandrova2008}
Alexandrova, O., Lacombe, C., \& Mangeney, A. 2008, Annales Geophysicae, 26,
  3585

\bibitem[{Bale {et~al.}(2005)Bale, Kellogg, Mozer, Horbury, \& Reme}]{Bale2005}
Bale, S.~D., Kellogg, P.~J., Mozer, F.~S., Horbury, T.~S., \& Reme, H. 2005,
  Physical Review Letters, 94, 1

\bibitem[{Birn \& Priest(2007)}]{birn2007reconnection}
Birn, J., \& Priest, E.~R. 2007, {Reconnection of magnetic fields:
  magnetohydrodynamics and collisionless theory and observations} (Cambridge
  University Press)

\bibitem[{Bruno \& Carbone(2013)}]{Bruno2013}
Bruno, R., \& Carbone, V. 2013, Living Reviews in Solar Physics, 10,
  doi:10.12942/lrsp-2013-2

\bibitem[{Burch {et~al.}(2016)Burch, Moore, Torbert, \& Giles}]{Burch2016}
Burch, J.~L., Moore, T.~E., Torbert, R.~B., \& Giles, B.~L. 2016, Space Science
  Reviews, 199, 5

\bibitem[{Capon(1969)}]{capon1969high}
Capon, J. 1969, Proceedings of the IEEE, 57, 1408

\bibitem[{Chen {et~al.}(2010)Chen, Horbury, Schekochihin, Wicks, Alexandrova,
  \& Mitchell}]{Chen2010}
Chen, C.~H., Horbury, T.~S., Schekochihin, A.~A., {et~al.} 2010, Physical
  Review Letters, 104, 1

\bibitem[{Chen \& Boldyrev(2017)}]{Chen2017}
Chen, C. H.~K., \& Boldyrev, S. 2017, The Astrophysical Journal, 842, 122

\bibitem[{Cho \& Lazarian(2004)}]{Cho_2004}
Cho, J., \& Lazarian, A. 2004, The Astrophysical Journal, 615, L41

\bibitem[{{Dai} {et~al.}(2020){Dai}, {Wang}, {Cai}, {Gonzalez}, {Hesse},
  {Escoubet}, {Phan}, {Vasyliunas}, {Lu}, {Li}, {Kong}, {Dunlop}, {Nakamura},
  {He}, {Fu}, {Zhou}, {Huang}, {Wang}, {Khotyaintsev}, {Graham}, {Retino},
  {Zelenyi}, {Grigorenko}, {Runov}, {Angelopoulos}, {Kepko}, {Hwang}, \&
  {Zhang}}]{Dai2020}
{Dai}, L., {Wang}, C., {Cai}, Z., {et~al.} 2020, Frontiers in Physics, 8, 89

\bibitem[{Duan {et~al.}(2020)Duan, He, Wu, \& Verscharen}]{Duan2020}
Duan, D., He, J., Wu, H., \& Verscharen, D. 2020, The Astrophysical Journal,
  896, 47

\bibitem[{Duan {et~al.}(2018)Duan, Pei, Huang, Wu, \& Verscharen}]{Duan2018}
Duan, D., Pei, Z., Huang, S., Wu, H., \& Verscharen, D. 2018, The Astrophysical
  Journal, 865, 89

\bibitem[{Ergun {et~al.}(2016)Ergun, Tucker, Westfall, Goodrich, Malaspina,
  Summers, Wallace, Karlsson, Mack, Brennan, Pyke, Withnell, Torbert, Macri,
  Rau, Dors, Needell, Lindqvist, Olsson, \& Cully}]{Ergun2016}
Ergun, R.~E., Tucker, S., Westfall, J., {et~al.} 2016, Space Science Reviews,
  199, 167

\bibitem[{Escoubet {et~al.}(2001)Escoubet, Fehringer, \&
  Goldstein}]{Escoubet2001}
Escoubet, C.~P., Fehringer, M., \& Goldstein, M. 2001, Annales Geophysicae, 19,
  1197

\bibitem[{Glassmeier {et~al.}(2001)Glassmeier, Motschmann, Dunlop, Balogh,
  Acu{\~{n}}a, Carr, Musmann, Forna{\c{c}}on, Schweda, Vogt, Georgescu, \&
  Buchert}]{Glassmeier2001}
Glassmeier, K.-H., Motschmann, U., Dunlop, M., {et~al.} 2001, Annales
  Geophysicae, doi:10.5194/angeo-19-1439-2001

\bibitem[{Goldreich \& Sridhar(1995)}]{Goldreich1995}
Goldreich, P., \& Sridhar, S. 1995, The Astrophysical Journal, 438, 763

\bibitem[{He {et~al.}(2011{\natexlab{a}})He, Marsch, Tu, Yao, \&
  Tian}]{He2011ICW}
He, J., Marsch, E., Tu, C., Yao, S., \& Tian, H. 2011{\natexlab{a}}, The
  Astrophysical Journal, 731, 85

\bibitem[{He {et~al.}(2011{\natexlab{b}})He, Tu, Marsch, \& Yao}]{He2012}
He, J., Tu, C., Marsch, E., \& Yao, S. 2011{\natexlab{b}}, The Astrophysical
  Journal Letters, 745, L8

\bibitem[{He {et~al.}(2015)He, Wang, Tu, Marsch, \& Zong}]{He2015}
He, J., Wang, L., Tu, C., Marsch, E., \& Zong, Q. 2015, Astrophysical Journal
  Letters, 800, L31

\bibitem[{{He} {et~al.}(2020){He}, {Zhu}, {Verscharen}, {Duan}, {Zhao}, \&
  {Wang}}]{2020ApJ...898...43H}
{He}, J., {Zhu}, X., {Verscharen}, D., {et~al.} 2020, \apj, 898, 43

\bibitem[{He {et~al.}(2019)He, Duan, Wang, Zhu, Li, Verscharen, Wang, Tu,
  Khotyaintsev, Le, \& Burch}]{He2019}
He, J., Duan, D., Wang, T., {et~al.} 2019, The Astrophysical Journal, 880, 121

\bibitem[{He {et~al.}(2011{\natexlab{c}})He, Marsch, Tu, Zong, Yao, \&
  Tian}]{He2011}
He, J.~S., Marsch, E., Tu, C.~Y., {et~al.} 2011{\natexlab{c}}, Journal of
  Geophysical Research: Space Physics, 116, 1

\bibitem[{{Hesse} {et~al.}(1999){Hesse}, {Schindler}, {Birn}, \&
  {Kuznetsova}}]{1999PhPl....6.1781H}
{Hesse}, M., {Schindler}, K., {Birn}, J., \& {Kuznetsova}, M. 1999, Physics of
  Plasmas, 6, 1781

\bibitem[{Horbury {et~al.}(2008)Horbury, Forman, \& Oughton}]{Horbury2008}
Horbury, T.~S., Forman, M., \& Oughton, S. 2008, Physical Review Letters, 101,
  1

\bibitem[{Horbury {et~al.}(2012)Horbury, Wicks, \& Chen}]{Horbury2012}
Horbury, T.~S., Wicks, R.~T., \& Chen, C.~H. 2012, Space Science Reviews, 172,
  325

\bibitem[{Howes {et~al.}(2011)Howes, Tenbarge, Dorland, Quataert, Schekochihin,
  Numata, \& Tatsuno}]{Howes2011}
Howes, G.~G., Tenbarge, J.~M., Dorland, W., {et~al.} 2011, Physical Review
  Letters, 107, 1

\bibitem[{Huang {et~al.}(2014)Huang, Sahraoui, Deng, He, Yuan, Zhou, Pang, \&
  Fu}]{Huang2014}
Huang, S.~Y., Sahraoui, F., Deng, X.~H., {et~al.} 2014, Astrophysical Journal
  Letters, 789, doi:10.1088/2041-8205/789/2/L28

\bibitem[{Leamon {et~al.}(1998)Leamon, Smith, Ness, Matthaeus, \&
  Wong}]{Leamon1998}
Leamon, R.~J., Smith, C.~W., Ness, N.~F., Matthaeus, W.~H., \& Wong, H.~K.
  1998, Journal of Geophysical Research: Space Physics, 103, 4775

\bibitem[{Lindqvist {et~al.}(2016)Lindqvist, Olsson, Torbert, King, Granoff,
  Rau, Needell, Turco, Dors, Beckman, Macri, Frost, Salwen, Eriksson,
  {\AA}hl{\'{e}}n, Khotyaintsev, Porter, Lappalainen, Ergun, Wermeer, \&
  Tucker}]{Lindqvist2016}
Lindqvist, P.~A., Olsson, G., Torbert, R.~B., {et~al.} 2016, Space Science
  Reviews, 199, 137

\bibitem[{Lu {et~al.}(2010)Lu, Huang, Xie, Wang, Wu, Vaivads, \& Wang}]{Lu2010}
Lu, Q., Huang, C., Xie, J., {et~al.} 2010, Journal of Geophysical Research:
  Space Physics, 115

\bibitem[{Matthaeus {et~al.}(1990)Matthaeus, Goldstein, \&
  Roberts}]{Matthaeus1990}
Matthaeus, W.~H., Goldstein, M.~L., \& Roberts, D.~A. 1990, Journal of
  Geophysical Research, 95, 20673

\bibitem[{Narita(2009)}]{Narita2009}
Narita, Y. 2009, 3031

\bibitem[{Narita(2018)}]{Narita2018}
---. 2018 (Springer International Publishing), 1--48,
  doi:10.1007/s41116-017-0010-0

\bibitem[{Narita {et~al.}(2010{\natexlab{a}})Narita, Sahraoui, Goldstein, \&
  Glassmeier}]{Narita2010MagneticEnergy}
Narita, Y., Sahraoui, F., Goldstein, M.~L., \& Glassmeier, K.~H.
  2010{\natexlab{a}}, Journal of Geophysical Research: Space Physics, 115, 1

\bibitem[{Narita {et~al.}(2010{\natexlab{b}})Narita, Sahraoui, Goldstein, \&
  Glassmeier}]{Narita2010doppler}
---. 2010{\natexlab{b}}, 115, 1

\bibitem[{Narita {et~al.}(2016)Narita, Plaschke, Nakamura, Baumjohann, Magnes,
  Fischer, V{\"{o}}r{\"{o}}s, Torbert, Russell, Strangeway, Leinweber, Bromund,
  Anderson, Le, Chutter, Slavin, Kepko, Burch, Motschmann, Richter, \&
  Glassmeier}]{Narita2016}
Narita, Y., Plaschke, F., Nakamura, R., {et~al.} 2016, 4774

\bibitem[{Oughton {et~al.}(2015)Oughton, Matthaeus, Wan, \&
  Osman}]{Oughton2015}
Oughton, S., Matthaeus, W.~H., Wan, M., \& Osman, K.~T. 2015, Philosophical
  Transactions of the Royal Society A: Mathematical, Physical and Engineering
  Sciences, 373, doi:10.1098/rsta.2014.0152

\bibitem[{Paschmann \& Schwartz(2000)}]{paschmann2000issi}
Paschmann, G., \& Schwartz, S. 2000, in Cluster-II Workshop
  Multiscale/Multipoint Plasma Measurements, Vol. 449, 99

\bibitem[{Pin{\c{c}}on \& Lefeuvre(1991)}]{pinccon1991local}
Pin{\c{c}}on, J.~L., \& Lefeuvre, F. 1991, Journal of Geophysical Research:
  Space Physics, 96, 1789

\bibitem[{Podesta(2012)}]{Podesta2012}
Podesta, J.~J. 2012, Journal of Geophysical Research: Space Physics, 117, 1

\bibitem[{Pollock {et~al.}(2016)Pollock, Moore, Jacques, Burch, Gliese, Saito,
  Omoto, Avanov, Barrie, Coffey, Dorelli, Gershman, Giles, Rosnack, Salo,
  Yokota, Adrian, Aoustin, Auletti, Aung, Bigio, Cao, Chandler, Chornay,
  Christian, Clark, Collinson, Corris, {De Los Santos}, Devlin, Diaz,
  Dickerson, Dickson, Diekmann, Diggs, Duncan, Figueroa-Vinas, Firman, Freeman,
  Galassi, Garcia, Goodhart, Guererro, Hageman, Hanley, Hemminger, Holland,
  Hutchins, James, Jones, Kreisler, Kujawski, Lavu, Lobell, LeCompte, Lukemire,
  MacDonald, Mariano, Mukai, Narayanan, Nguyan, Onizuka, Paterson, Persyn,
  Piepgrass, Cheney, Rager, Raghuram, Ramil, Reichenthal, Rodriguez, Rouzaud,
  Rucker, Saito, Samara, Sauvaud, Schuster, Shappirio, Shelton, Sher, Smith,
  Smith, Smith, Steinfeld, Szymkiewicz, Tanimoto, Taylor, Tucker, Tull, Uhl,
  Vloet, Walpole, Weidner, White, Winkert, Yeh, \& Zeuch}]{Pollock2016}
Pollock, C., Moore, T., Jacques, A., {et~al.} 2016, Space Science Reviews, 199,
  331

\bibitem[{Robert {et~al.}(1998)Robert, Roux, Harvey, Dunlop, Daly, \&
  Glassmeier}]{robert1998tetrahedron}
Robert, P., Roux, A., Harvey, C.~C., {et~al.} 1998, Analysis methods for
  multi-spacecraft data, 323

\bibitem[{Roberts {et~al.}(2019)Roberts, Narita, Nakamura, V{\"{o}}r{\"{o}}s,
  \& Gershman}]{Roberts2019}
Roberts, O.~W., Narita, Y., Nakamura, R., V{\"{o}}r{\"{o}}s, Z., \& Gershman,
  D. 2019, Frontiers in Physics, 7, 1

\bibitem[{Russell {et~al.}(2016)Russell, Anderson, Baumjohann, Bromund,
  Dearborn, Fischer, Le, Leinweber, Leneman, Magnes, Means, Moldwin, Nakamura,
  Pierce, Plaschke, Rowe, Slavin, Strangeway, Torbert, Hagen, Jernej,
  Valavanoglou, \& Richter}]{Russel2016}
Russell, C.~T., Anderson, B.~J., Baumjohann, W., {et~al.} 2016, Space Science
  Reviews, 199, 189

\bibitem[{Sahraoui {et~al.}(2010{\natexlab{a}})Sahraoui, Belmont, Goldstein, \&
  Rezeau}]{Sahraoui2010limitation}
Sahraoui, F., Belmont, G., Goldstein, M.~L., \& Rezeau, L. 2010{\natexlab{a}},
  Journal of Geophysical Research A: Space Physics, 115, 1

\bibitem[{Sahraoui {et~al.}(2006)Sahraoui, Belmont, Rezeau, Cornilleau-Wehrlin,
  Pin{\c{c}}on, \& Balogh}]{Sahraoui2006}
Sahraoui, F., Belmont, G., Rezeau, L., {et~al.} 2006, Physical Review Letters,
  96, 1

\bibitem[{Sahraoui {et~al.}(2010{\natexlab{b}})Sahraoui, Goldstein, Belmont,
  Canu, \& Rezeau}]{Sahraoui2010}
Sahraoui, F., Goldstein, M.~L., Belmont, G., Canu, P., \& Rezeau, L.
  2010{\natexlab{b}}, Physical Review Letters, 105, 1

\bibitem[{Sahraoui {et~al.}(2020)Sahraoui, Hadid, \& Huang}]{Sahraoui2020}
Sahraoui, F., Hadid, L., \& Huang, S. 2020 (Springer Singapore), 1--33,
  doi:10.1007/s41614-020-0040-2

\bibitem[{Sahraoui {et~al.}(2003)Sahraoui, Pin{\c{c}}on, Belmont, Rezeau,
  Cornilleau-Wehrlin, Robert, Mellul, Bosqued, Balogh, Canu, \&
  Chanteur}]{Sahraoui2003}
Sahraoui, F., Pin{\c{c}}on, J.~L., Belmont, G., {et~al.} 2003, Journal of
  Geophysical Research: Space Physics, 108, 1

\bibitem[{Schekochihin {et~al.}(2009)Schekochihin, Cowley, Dorland, Hammett,
  Howes, Quataert, \& Tatsuno}]{Schekochihin2009}
Schekochihin, A.~A., Cowley, S.~C., Dorland, W., {et~al.} 2009, Astrophysical
  Journal, Supplement Series, 182, 310

\bibitem[{Tjulin {et~al.}(2005)Tjulin, Pin{\c{c}}on, Sahraoui, Andr{\'{e}}, \&
  Cornilleau-Wehrlin}]{Tjulin2005}
Tjulin, A., Pin{\c{c}}on, J.~L., Sahraoui, F., Andr{\'{e}}, M., \&
  Cornilleau-Wehrlin, N. 2005, Journal of Geophysical Research: Space Physics,
  110, 1

\bibitem[{Verscharen \& Chandran(2018)}]{Verscharen2018NHDS}
Verscharen, D., \& Chandran, B. D.~G. 2018, Research Notes of the {AAS}, 2, 13

\bibitem[{Wang {et~al.}(2020)Wang, He, Alexandrova, Dunlop, \&
  Perrone}]{Wang2020}
Wang, T., He, J., Alexandrova, O., Dunlop, M., \& Perrone, D. 2020

\bibitem[{Wu {et~al.}(2020)Wu, Tu, Wang, He, Yang, \& Wang}]{Wu2020}
Wu, H., Tu, C., Wang, X., {et~al.} 2020, The Astrophysical Journal, 892, 138

\bibitem[{Zenitani {et~al.}(2011)Zenitani, Hesse, Klimas, \&
  Kuznetsova}]{Zenitani2011}
Zenitani, S., Hesse, M., Klimas, A., \& Kuznetsova, M. 2011, Physical Review
  Letters, 106, 1

\bibitem[{Zhang {et~al.}(2020)Zhang, He, Narita, \& Feng}]{zhang2020conditions}
Zhang, L., He, J.~S., Narita, Y., \& Feng, X.~S. 2020, On the Conditions of
  Aliasing Effects in Constellations with More than Four Spacecraft, , ,
  arXiv:2011.05672

\bibitem[{Zhu {et~al.}(2019)Zhu, He, Verscharen, \& Zhao}]{Zhu_2019}
Zhu, X., He, J., Verscharen, D., \& Zhao, J. 2019, The Astrophysical Journal,
  878, 48

\end{thebibliography}

\end{document}